\documentclass[conference,a4paper]{IEEEtran}

\usepackage[utf8]{inputenc} 
\usepackage[T1]{fontenc}
\usepackage{url}     
\usepackage{cite}
\usepackage{amsmath,amssymb,placeins}
\usepackage{threeparttable}
\usepackage{algorithmic}
\usepackage{graphicx}
\usepackage{textcomp}
\usepackage[mathscr]{euscript}
\usepackage{lipsum}
\usepackage{array}
\usepackage{cuted}
\usepackage{mathtools}
\usepackage{multicol}
\usepackage{array}
\usepackage{makecell}
\usepackage{diagbox}
\usepackage[symbol]{footmisc}
\usepackage{caption}
\usepackage{tabularx}

\usepackage[mathscr]{euscript}
\usepackage[usenames]{color}
\usepackage{amsfonts}
\usepackage{latexsym}
\usepackage{subfigure}
\usepackage[format=hang]{subfig}
\usepackage{caption}
\usepackage{floatrow}
\usepackage{multirow}
\usepackage{epstopdf}

\newtheorem{theorem}{{{\textit{Theorem}}}}

\newtheorem{lemma}{{{\textit{Lemma}}}}

\newtheorem{definition}{{{\textit{Definition}}}}

\newtheorem{remark}{{{\textit{Remark}}}}

\newtheorem{example}{{{\textit{Example}}}}

\newcommand{\Mod}[1]{\ (\mathrm{mod}\ #1)}
\usepackage{cite}             

\usepackage{mleftright}       
\mleftright                   

\usepackage{graphicx}         
\usepackage{booktabs}         

\begin{document}

\title{A Direct Construction of Optimal Symmetrical Z-Complementary Code Sets of Prime Power Lengths} 


%
%
%
\author{%
   \IEEEauthorblockN{Praveen Kumar \IEEEauthorrefmark{1},
                     Sudhan Majhi\IEEEauthorrefmark{2},
                    Subhabrata Paul\IEEEauthorrefmark{1}}
  \IEEEauthorblockA{\IEEEauthorrefmark{1}%
                     Department of Mathematics, IIT Patna, India
                    \{praveen\_2021ma03, subhabrata\}@iitp.ac.in}
  \IEEEauthorblockA{\IEEEauthorrefmark{2}%
                   Department of Electrical Communication Engineering,
IISC Bangalore, India,
                     smajhi@iisc.ac.in}
 }

\maketitle

\begin{abstract}
  This paper presents a direct construction of an optimal symmetrical Z-complementary code set (SZCCS) of prime power lengths using a multi-variable function (MVF). SZCCS is a natural extension of the Z-complementary code set (ZCCS), which has only front-end zero correlation zone (ZCZ) width. SZCCS has both front-end and tail-end ZCZ width. SZCCSs are used in developing optimal training sequences for broadband generalized spatial modulation systems over frequency-selective channels because they have ZCZ width on both the front and tail ends. The construction of optimal SZCCS with large set sizes and
prime power lengths is presented for the first time in this paper. Furthermore, it is worth noting that several existing works on ZCCS and SZCCS can be viewed as special cases of the proposed construction.
\end{abstract}
\section{Introduction}\label{sec:intro}
{T}{he} idea of Z-complementary pairs (ZCPs) was introduced by Fan \textit{et al.} \cite{fan2007}. The sum of the aperiodic auto-correlation function (AACF) of the two sequences in a ZCP is zero within a particular zone, which is referred to as the zero-correlation zone (ZCZ). When this ZCZ width $Z$ is equal to the sequence length $N$, ZCP becomes a Golay complementary pair (GCP). Unlike GCPs, ZCPs are available in arbitrary lengths with various ZCZ widths \cite{zcp1,zcp2,zcp3,zcp4,praveen1}.

The idea of ZCPs introduced in \cite{fan2007} was generalized to a Z-complementary code set (ZCCS)
by Feng \textit{et al.} in \cite{zccs1}.
ZCCS only considers the front-end ZCZ of the AACFs and aperiodic cross-correlation functions (ACCFs). Several generalized Boolean functions (GBFs) based constructions of ZCCSs of non-power-of-two lengths are proposed in the literature\cite{sahin,pmz,zhou4,gobinda,chunlei1,yubo3}. A ZCCS with $K$ codes, with each code having $M$ sequences each of length $N$ and ZCZ width $Z$ is denoted by $(K,M,N,Z)$-ZCCS. For the special case, $K=M$ and $N=Z$, it is known as a complete complementary code (CCC) and is denoted by $(K,K,N)$-CCC. Recently, Li \textit{et al.} proposed a direct construction of multiple CCC of prime power length and with inter set ZCZ width using a multi-variable function (MVF) \cite{multiple_ccc_2022}. After combining these multiple CCC, optimal $(p^{n+v},p^n,p^m,p^{m-v})$-ZCCS are obtained \cite{multiple_ccc_2022}. 

Recently, the idea of ZCCS has been extended to symmetrical-ZCCS (SZCCS), which exhibits ZCZ properties for the front-end and the tail-end \cite{szccs1}. The authors in \cite{szccs1} have presented a GBF based construction of $(8,2,2^m,2^{m-2}-1)$-SZCCS. In practice, a front-end ZCZ and a tail-end ZCZ have a particular role in mitigating interference with small and large delays, respectively.
SZCCSs with larger set sizes are used in designing training sequences for broadband generalized spatial modulation (GSM) systems over frequency-selective channels \cite{szccs1}. The fact that SZCCS is being widely applied in the GSM systems and also the unavailability of constructions with flexible parameters in terms of set size, flock size, and length have inspired the authors to propose a direct construction of SZCCSs of prime power lengths in this paper. 

 The proposed MVF-based construction of SZCCS has a set size $p^{k+\delta}$, which is much larger than the flock size of $p^k,$ where $m\geq 3,~0\leq\delta<m,~1\leq k \leq m-\delta.$ Also, the proposed SZCCS has large ZCZ width of $p^{m-\delta}-1$, and it achieves the optimality condition. 
Many of the existing constructions of ZCCS and SZCCS appear as special cases of the proposed construction.

The rest of the paper is structured as follows: preliminary work is covered in Section II, and the proposed SZCCS construction based on MVF is presented in Section III. Section IV of the paper presents a comprehensive comparison between the proposed constructions and the existing works, providing detailed insights. Following that, Section V concludes the paper.

\section{Preliminaries}
The essential concepts, notations, and previously established findings necessary for the proposed construction are explained in this section.
\begin{definition}\label{def1} Let $\mathbf{u}$ $=(u_{0},u_{1}, \ldots, u_{N-1})$ and $\mathbf{v}$ $=(v_{0},v_{1}, \ldots, v_{N-1})$ be two sequences of length $N$ over $\mathbb{Z}_{q}$. At a shift $\tau$, the  ACCF is defined as
\begin{equation}\label{eqn5.1}
\mathcal{C}\left({\mathbf{u}, \mathbf{v}}\right)(\tau)=\begin{cases}
\sum_{i=0}^{N-1-\tau} \omega_q^{u_{i}-v_{i+\tau}}, & 0 \leq \tau \leq N-1, \\
\sum_{i=0}^{N-1+\tau} \omega_q^{u_{i-\tau}-v_{i}}, & -N+1 \leq \tau \leq-1, \\
0, & |\tau| \geq N,
\end{cases}
\end{equation}
 where $q$ is a positive integer greater than $2$, and $\omega_q=\exp(2\pi\sqrt{-1}/q)$. For the special case, $\mathbf{u}=\mathbf{v},~ \mathcal{C}(\mathbf{u}, \mathbf{v})(\tau)$ is referred to as the AACF of $\mathbf{u}$ and is denoted by $\mathcal{A}(\mathbf{u})(\tau)$.
 \end{definition}
 Consider a set  $\mathrm{C}=\left\{{C^{0}}, {C^{1}}, \hdots, {C^{K-1}}\right\}$, where each set ${C^{e}}$ consists of $M$ sequences, i.e., ${C^{e}}=\left\{\mathbf{c}_{0}^{e}, \mathbf{c}_{1}^{e}, \hdots, \mathbf{c}_{M-1}^{e}\right\}$, and length of each sequence $\mathbf{c}_{l}^{e}$ is $N$, where $0 \leq e \leq K-1$ and $0 \leq l \leq M-1$.
 \begin{definition}\label{def2}
  The set $\mathrm{C}$ defined above is called a ZCCS, denoted by $(K,M,N,Z)$-ZCCS, if the ACCF of $C^{e}$ and $C^{e^{\prime}}$ satisfies
\begin{equation}\label{eqn5.2}
\begin{aligned}
\mathcal{C}\left(C^{e}, C^{e^{\prime}}\right)(\tau)&=\sum_{l=0}^{M-1} \mathcal{C}\left(\mathbf{c}_{l}^{e}, \mathbf{c}_{l}^{e^{\prime}}\right)(\tau) \\&
= \begin{cases}MN, & \tau=0, e=e^{\prime}, \\ 0, & 0 < |\tau| < Z, e=e', \\ 0, &  |\tau| < Z, e\neq e', \end{cases}
\end{aligned}
\end{equation}
where $0 \leq e, e' \leq K-1$. For a general $(K,M,N,Z)$-ZCCS, $K\leq M \lfloor N/Z \rfloor$, whereas for the special case, $K= M \lfloor N/Z \rfloor$, it becomes  an optimal ZCCS \cite{zccs_bound}. 
 Again, for the special case, $K=M$ and $Z=N$, the ZCCS is called a CCC of order $K$ and length $N$, and is denoted by $(K,K,N)$-CCC.
 \end{definition}
\begin{definition}\label{DEF3}
   The set $\mathrm{C}$ defined above is known as a SZCSS, denoted by $(K,M,N,Z)$-SZCCS, if for $\mathcal{T}_{1}=\{1,2, \cdots, Z\}$ and $\mathcal{T}_{2}=\{N-Z, N-$ $Z+1, \cdots, N-1\}$ with $Z \leq N,$ it satisfies the following properties
\begin{equation}\label{eqn5.3}
    \begin{aligned}
      &P1: \sum_{i=0}^{M-1} \mathcal{A}(\mathbf{c}_i^e)(\tau)=0, \quad \text { for all }|\tau| \in \left(\mathcal{T}_{1} \cup \mathcal{T}_{2}\right)\cap \mathcal{T};  \\&
      P2: \sum_{i=0}^{M-1} \mathcal{C}(\mathbf{c}_i^e, \mathbf{c}_i^{e'})(\tau)=0, \quad \text{ for all } |\tau| \in\{0\} \cup \mathcal{T}_{1} \cup \mathcal{T}_{2};\\&
    \end{aligned}
\end{equation}
where $\mathcal{T}=\{1,2,\hdots,N-1\}$ and $e \neq e'$. For a $(K,M,N,Z)$-SZCCS, $K,M,N$ and $Z$ are known as the set size, flock size, sequence length, and ZCZ width, respectively.
\end{definition}

For a MVF $f:\{0,1,\hdots,p-1\}^m \rightarrow \mathbb{Z}_q$ in $m$ $p$-ary variables $x_1,x_2,\hdots,x_{m}$, corresponding $\mathbb{Z}_q$-valued sequence is denoted by $\mathbf{f}$, and calculated as
$\mathbf{f}=\left({f_0}, {f_1}, \hdots,{f_{p^m-1}}\right)$ ,
where $f_r=f({r_1},{r_2},\hdots,r_{m})$, and $r=({r_1},{r_2},\hdots,r_{m})$ is the $p$-ary representation of the integer $r$, i.e., $r=\sum_{i=1}^{m}r_ip^{i-1}$. The complex-valued sequence associated with $\mathbf{f}$ is produced as $\Psi(f)=$ $\left(\omega_q^{f_0},\omega_q^{f_1},\hdots,\omega_q^{f_{p^m-1}}\right)$, where $\omega_q$ is the $q$-th $(q\geq2)$ root of unity \cite{multiple_ccc_2022}.

\begin{lemma}[\cite{szccs1,zccs_bound}]{\label{lem5.1}} Any unimodular $(K,M, N,Z)$-SZCCS satisfies $K \leq M\left\lfloor\frac{N}{Z+1}\right\rfloor$, when $K =M\left\lfloor\frac{N}{Z+1}\right\rfloor$, it is known as an optimal $(K,M,N,Z)$-SZCCS.
\end{lemma}
\section{Proposed Construction}
This section explains a MVF based construction of optimal $(p^{k+\delta},p^k,p^m,p^{m-\delta}-1)$-SZCCS.
\begin{theorem}\label{thm5.1}
For any positive integer $m \geq 3$ and $0\leq\delta <m$, we let the set $\{1,2,\hdots,m-\delta\}$ be partitioned into $k$ sets, $S_1,S_2,\hdots,S_k$, where $1\leq k\leq m-\delta$. Let the cardinality of the set $S_\beta$ be $m_\beta$, and $\pi_\beta$ be a one-one and onto mapping from $\{1,2,\hdots,m_\beta\}$ to $S_\beta$ for $\beta=1,2,\hdots,k.$ Let us define a MVF $g:\{0,1,\hdots,p-1\}^m \rightarrow \mathbb{Z}_q$ by
\begin{equation}\label{eqn5.4}
    g= \frac{q}{p} \sum_{\beta=1}^{k} \sum_{\gamma=1}^{m_{\beta}-1} x_{\pi_{\beta}(\gamma)} x_{\pi_{\beta}(\gamma+1)}+\sum_{l=1}^{m}\lambda_lx_l +\lambda_0,
\end{equation}
where $p$ is a prime and $p\mid q$, $\lambda_l \in \mathbb{Z}_q$, $0\leq l \leq m$.
 Then the set ${A}=\left\{A^0,A^1,\hdots,A^{p^{k+\delta}-1}\right\}$ is a $(p^{k+\delta},p^k,p^m,p^{m-\delta}-1)$-SZCCS, where $A^t=
 \{\mathbf{a}^t_0,\mathbf{a}^t_1,\hdots,\mathbf{a}^t_{p^{k}-1}\}$ and 
\begin{equation}
\begin{aligned}
    \mathbf{a}^t_\sigma=\boldsymbol {g}&+\frac{q}{p}\left(\sum_{\beta=1}^{k}\sigma_\beta \boldsymbol{x}_{\pi_\beta(1)}+\sum_{\beta=1}^{k}t_\beta \boldsymbol{x}_{\pi_\beta(m_{\beta})}\right.\\& \left.+\sum_{\beta=1}^{\delta}t_{k+\beta} \boldsymbol{x}_{m-\delta+\beta}\right),
    \end{aligned}
\end{equation}
for $\sigma=0,1,\hdots,p^k-1$ and $t=0,1,\hdots,p^{k+\delta}-1$ with $p$-ary vector representation $(\sigma_1,\sigma_2,\hdots,\sigma_k)$ and $(t_1,t_2,\hdots,t_k,\hdots,t_{k+\delta})$, respectively.
\end{theorem}
\begin{IEEEproof}
 We will prove that the set ${A}=\left\{A^0,A^1,\hdots,A^{p^{k+\delta}-1}\right\}$ obtained from \textit{Theorem} \ref{thm5.1} satisfies two conditions $P1$ and $P2$ given in \textit{Definition} \ref{DEF3},  with $K=p^{k+\delta}, M=p^{k},N=p^{m},$ and $Z=p^{m-\delta}-1$, and hence a $(p^{k+\delta},p^k,p^{m},p^{m-\delta}-1)$-SZCCS. Let the $p$-ary representations of  $0 \leq r,s<p^m$ be $(r_1, r_2,\hdots,r_m)$ and $(s_1, s_2,\hdots,s_m)$, respectively. Let $\mathbf{a}^t_{\sigma}=(a^t_{\sigma,0},a^t_{\sigma,1},\hdots,a^t_{\sigma,p^m-1})$.

First, we prove the property $P1$ of \textit{Definition} \ref{DEF3}, i.e., for $\mathcal{T}_1=\{1,2,\hdots,p^{m-\delta}-1\}$ and $\mathcal{T}_2=\{p^m-p^{m-\delta}-1,p^m-p^{m-\delta},\hdots,p^m-1\}$ the AACF for every $A^t$ zero, i.e.,
\begin{equation}\label{eqn5.6}
  \sum_{\sigma=0}^{p^k-1} \mathcal{A}(\mathbf{a}_\sigma^t)(\tau)=\sum_{r=0}^{p^{m}-\tau-1}\sum_{\sigma=0}^{p^{k}-1}\omega_q^{\left(a^t_{\sigma,r}-a^t_{\sigma,r+\tau}\right)}=0,
\end{equation}
  for all $|\tau| \in \left(\mathcal{T}_{1} \cup \mathcal{T}_{2}\right)\cap \mathcal{T},$ where $\mathcal{T}=\{1,2,\hdots,p^m-1\}$.
Let $s = r + \tau$ for any integer $r$. 
Then we consider the two cases listed below.

\textit{Case I:} $r_{\pi_{\beta}(1)}\neq s_{\pi_\beta(1)}$ for some $ \beta \in \{1,2,\hdots,k\}$.
Then there exist $\mathbf{a}^t_{\sigma^j}=(a^t_{\sigma^j,0},a^t_{\sigma^j,1},\hdots,a^t_{\sigma^j,p^m-1})=\mathbf{a}^t_\sigma+(jq/p)\boldsymbol{x}_{\pi_\beta(1)} \in A^t$, where $1\leq j \leq p-1$, such that 
\begin{equation}\label{eqn5.7}
  {a^t_{\sigma^j, r}-a^t_{\sigma, r}}=\frac{jq}{p} {r}_{\pi_\beta(1)},
\end{equation}
and
\begin{equation}\label{eqn5.8}
  {a^t_{\sigma^j, s}-a^t_{\sigma, s}}=\frac{jq}{p} {s}_{\pi_\beta(1)}.
\end{equation}
So from the above two equations, we get 
\begin{equation}\label{eqn5.9}
\left(a^t_{\sigma^j, r}-a^t_{\sigma^j, s}\right)-\left(a^t_{\sigma, r}-
  a^t_{\sigma, s} \right)=\frac{jq}{p}\left({r}_{\pi_\beta(1)}-{s}_{\pi_\beta(1)} \right).
  \end{equation}
 Raising to the power of $q$-th root of unity and then taking sum over $1 \leq j \leq p-1$, we get
  \begin{equation}\label{eqn5.10}
\sum_{j=1}^{p-1}  \omega_q^{\left(a^t_{\sigma^j, r}-a^t_{\sigma^j, s}\right)-\left(a^t_{\sigma, r}-a^t_{\sigma, s} \right)} =\sum_{j=1}^{p-1}\omega_p^{j\left({r}_{\pi_\beta(1)}-{s}_{\pi_\beta(1)} \right)}=-1.
\end{equation}
Hence,
\begin{equation}\label{eqn5.11}
    \omega_q^{\left(a^t_{\sigma, r}-a^t_{\sigma, s} \right)}+\sum_{j=1}^{p-1}  \omega_q^{\left(a^t_{\sigma^j, r}-a^t_{\sigma^j, s}\right)}=0.
\end{equation}
So, the sum AACF is zero. i.e.,
\begin{equation}\label{eqn5.12}
\sum_{\sigma=0}^{p^k-1} \mathcal{A}(\mathbf{a}_\sigma^t)(\tau)=0.
\end{equation}
\textit{Case II:}  $r_{\pi_\beta(1)}=s_{\pi_\beta(1)}$ for all $\beta\in\{1,2, \ldots, k\}$. Then, there exist integers $\hat{\beta}$ and $\hat{\gamma}$ such that  $\hat{\beta}$ is the largest integer satisfying $r_{\pi_\beta(\gamma)}=s_{\pi_\beta(\gamma)}$ for all $\beta=1,2,\hdots,\hat{\beta}-1$, and $\gamma=1,2,\hdots,m_{\alpha}$, and $\hat{\gamma}$ is the least integer satisfying $r_{\pi_{\hat{\beta}}(\hat{\gamma})}\neq s_{\pi_{\hat{\beta}}(\hat{\gamma})}$ . If the above statement doesn't hold, then we have $r_i=s_i$ for $i=$ $1,2, \ldots, m-\delta$ since $\bigcup_{\alpha=1}^k I_\alpha=\{1,2, \ldots, m-\delta\}$. Hence the lower bound of $\tau$ is estimated as,
\begin{equation}\label{eqn5.13}
 \tau=s-r=\sum_{i=m-\delta+1}^m\left(s_i-r_i\right) p^{i-1} \geq p^{m-\delta},   
\end{equation}
also the upper bound of $\tau$ is estimated as,
\begin{equation}\label{eqn5.14}
\begin{aligned}
 \tau=&s-r=\sum_{i=m-\delta+1}^m\left(s_i-r_i\right) p^{i-1} \\& \leq p^{m-\delta}+p^{m-\delta+1}+\hdots+p^{m-1}=p^{m}-p^{m-\delta}.
 \end{aligned}
\end{equation}
The above two equations (\ref{eqn5.13}) and (\ref{eqn5.14}) contradict the assumption that $\tau \in \left(\mathcal{T}_{1} \cup \mathcal{T}_{2}\right)\cap \mathcal{T}$. So this guarantees the existence of integers $\hat{\beta}$ and $\hat{\gamma}$ with the above-mentioned conditions.  Let $r^j$ and $s^j$ be  two integer which differs from $r$ and $s$, respectively, in only one position $\pi_{\hat{\beta}}(\hat{\gamma}-1)$ of their $p$-ary representation, i.e., $r^j_{\pi_{\hat{\beta}}(\hat{\gamma}-1)}=r_{\pi_{\hat{\beta}}(\hat{\gamma}-1)}-j$ and $s^j_{\pi_{\hat{\beta}}(\hat{\gamma}-1)}=s_{\pi_{\hat{\beta}}(\hat{\gamma}-1)}-j$, where $1 \leq j \leq p-1$. Since $s=r+\tau$, we get $s^j=r^j+\tau$. The difference between the terms $a^t_{\sigma,r}-a^t_{\sigma,r^j}$ is calculated below as
\begin{equation}\label{eqn5.15}
    \begin{aligned}
    a^t_{\sigma,r}-a^t_{\sigma,r^j}&=g_r-g_{r^j}\\&=j\left(\frac{q}{p}r_{\pi_{\hat{\beta}}(\hat{\gamma}-2)}+\frac{q}{p}r_{\pi_{\hat{\beta}}(\hat{\gamma})}+\lambda_{\pi_{\hat{\beta}}(\hat{\gamma}-1)} \right).
    \end{aligned}
\end{equation}
Similarly, it can be calculated that
\begin{equation}\label{eqn5.16}
    a^t_{\sigma,s}-a^t_{\sigma,s^j}=j\left(\frac{q}{p}s_{\pi_{\hat{\beta}}(\hat{\gamma}-2)}+\frac{q}{p}s_{\pi_{\hat{\beta}}(\hat{\gamma})}+\lambda_{\pi_{\hat{\beta}}(\hat{\gamma}-1)} \right).
\end{equation}
From the above two equations and using $r_{\pi_{\hat{\beta}}(\hat{\gamma}-2)}\neq s_{\pi_{\hat{\beta}}(\hat{\gamma}-2)}$, we get the following equality
\begin{equation}\label{eqn5.17}
   a^t_{\sigma,r^j}-a^t_{\sigma,s^j}-\left( a^t_{\sigma,r}-a^t_{\sigma,s} \right)=j\frac{q}{p}\left(s_{\pi_{\hat{\beta}}(\hat{\gamma})}-r_{\pi_{\hat{\beta}}(\hat{\gamma})} \right).
\end{equation}
 Raising to the power of $q$-th root of unity and then taking sum over $1 \leq j \leq p-1$, we get the following expression 
\begin{equation}\label{eqn5.18}
\sum_{j=1}^{p-1}  \omega_q^{\left(a^t_{\sigma, r^j}-a^t_{\sigma, s^j}\right)-\left(a^t_{\sigma, r}-a^t_{\sigma, s} \right)} =\sum_{j=1}^{p-1}\omega_p^{j\left(s_{\pi_{\hat{\beta}}(\hat{\gamma})}-r_{\pi_{\hat{\beta}}(\hat{\gamma})} \right)}=-1.
\end{equation}
Hence,
\begin{equation}\label{eqn5.19}
    \omega_q^{\left(a^t_{\sigma, r}-a^t_{\sigma, s} \right)}+\sum_{j=1}^{p-1}  \omega_q^{\left(a^t_{\sigma, r^j}-a^t_{\sigma, s^j}\right)}=0.
\end{equation}
So, the AACF is zero.
Combining \textit{Cases} I and II, we get AACF of $A^t$ is zero for $\tau \in \left(\mathcal{T}_{1} \cup \mathcal{T}_{2}\right)\cap \mathcal{T}$.

Next, in the following part, we will demonstrate that any two distinct sets $A^{t_1}$ and $A^{t_2}$ have zero ACCF for all $|\tau| \in\{0\} \cup \mathcal{T}_{1} \cup \mathcal{T}_{2}$, i.e.,
\begin{equation}\label{eqn5.20}
  \sum_{\sigma=0}^{p^k-1} \mathcal{C}(\mathbf{a}_\sigma^{t_1},\mathbf{a}_\sigma^{t_2})(\tau)=\sum_{r=0}^{p^{m}-\tau-1}\sum_{\sigma=0}^{p^{k}-1}\omega_q^{\left(a^{t_1}_{\sigma,r}-a^{t_2}_{\sigma,r+\tau}\right)}=0.
\end{equation}
Similar to the first part, we let $s=r+\tau$ for any integer $r$ and consider two cases.

\textit{Case I}: Suppose $r_{\pi_\beta(1)} \neq s_{\pi_\beta(1)}$ for some $\beta \in\{1,2, \ldots, k\}$. In the same manner as \textit{Case I} in the first part there exist $\mathbf{a}^t_{\sigma^j}=(a^t_{\sigma^j,0},a^t_{\sigma^j,1},\hdots,a^t_{\sigma^j,p^m-1})=\mathbf{a}^t_\sigma+(jq/p)\boldsymbol{x}_{\pi_\beta(1)} \in A^t$, for $t=t_1,t_2$, where $1\leq j \leq p-1$, such that
\begin{equation}\label{eqn5.21}
\sum_{j=1}^{p-1}  \omega_q^{\left(a^{t_1}_{\sigma^j, r}-a^{t_2}_{\sigma^j, s}\right)-\left(a^{t_1}_{\sigma, r}-a^{t_2}_{\sigma, s} \right)} =\sum_{j=1}^{p-1}\omega_p^{j\left({r}_{\pi_\beta(1)}-{s}_{\pi_\beta(1)} \right)}=-1.
\end{equation}
Hence, similar to the first part, the ACCF becomes zero, i.e.,
\begin{equation}\label{eqn5.22}
 \sum_{\sigma=0}^{p^{k}-1}\omega_q^{\left(a^{t_1}_{\sigma,r}-a^{t_2}_{\sigma,s}\right)}=0.
\end{equation}
\textit{Case II:} Suppose we have $r_{\pi_\beta(1)}=s_{\pi_\beta(1)}$ for all $\beta \in \{1,2, \ldots, k\}$. As argued in \textit{Case II} in the first part, a similar result can be obtained, i.e.,
\begin{equation}\label{eqn5.23}
    \omega_q^{\left(a^{t_1}_{\sigma, r}-a^{t_2}_{\sigma, s} \right)}+\sum_{j=1}^{p-1}  \omega_q^{\left(a^{t_1}_{\sigma, r^j}-a^{t_2}_{\sigma, s^j}\right)}=0.
\end{equation}
So, from \textit{Case I} and \textit{Case II}, we get $\sum_{\sigma=0}^{p^k-1} \mathcal{C}(\mathbf{a}_\sigma^{t_1},\mathbf{a}_\sigma^{t_2})(\tau)=0$, for $|\tau| \in \mathcal{T}_{1} \cup \mathcal{T}_{2}$.
 It only suffices to show that
\begin{equation}\label{eqn5.24}
  \sum_{\sigma=0}^{p^k-1} \mathcal{C}(\mathbf{a}_\sigma^{t_1},\mathbf{a}_\sigma^{t_2})(0)=\sum_{\sigma=0}^{p^{k}-1}\sum_{r=0}^{p^{m}-1}\omega_q^{\left(a^{t_1}_{\sigma,r}-a^{t_2}_{\sigma,r}\right)}=0.
\end{equation}
Let $\oplus$ denotes modulo-$p$ addition; let $\left(t_{11}, t_{12}, \ldots, t_{1k+\delta}\right)$ and $\left(t_{21}, t_{22}, \ldots, t_{2k+\delta}\right)$ denote the $p$-ary representations of $t_1$ and $t_2$, respectively. Then, for $\sigma=0,1, \hdots, p^k-1$, we have $\mathbf{a}_\sigma^{t_1}-\mathbf{a}_\sigma^{t_2} \equiv(q / p) \mathbf{d}\Mod q \quad$ where $\quad \mathbf{d}=$ $\left(t_{11} \oplus t_{21}\right) \boldsymbol{x}_{\pi_1\left(m_1\right)} \oplus\left(t_{12} \oplus t_{22}\right) \boldsymbol{x}_{\pi_2\left(m_2\right)} \oplus \cdots \oplus\left(t_{1k} \oplus t_{2k}\right) \boldsymbol{x}_{\pi_k\left(m_k\right)} \oplus$ $\left(t_{1k+1} \oplus t_{1k+1}\right) \boldsymbol{x}_{m-\delta+1} \oplus \cdots \oplus\left(t_{1k+\delta} \oplus t_{2k+\delta}\right) \boldsymbol{x}_m$. For every $1 \leq \beta \leq k$, $\boldsymbol{x}_{\pi_{\beta}\left(m_\beta\right)}$ takes each value from $\{0,1,\hdots,p-1\}$ exactly $p^{m-1}$ times.
Also for $1 \leq \beta \leq \delta$, $\boldsymbol{x}_{m-\delta+\beta}$ takes each value from $\{0,1,\hdots,p-1\}$ exactly $p^{m-1}$ times. So, $\mathbf{d}$ also takes each value from  $\{0,1,\hdots,p-1\}$ exactly $p^{m-1}$ times, i.e., it is balanced.
So, 
\begin{equation}\label{eqn5.25}
  \sum_{r=0}^{p^{m}-1}\omega_q^{\left(a^{t_1}_{\sigma,r}-a^{t_2}_{\sigma,r}\right)}= \sum_{r=0}^{p^{m}-1}\omega_p^{d_r}=0.
\end{equation}
Therefore, from (\ref{eqn5.24}) and (\ref{eqn5.25}) the ACCF at shift $\tau=0$ is zero, i.e.,
\begin{equation}\label{eqn5.26}
  \sum_{\sigma=0}^{p^k-1} \mathcal{C}(\mathbf{a}_\sigma^{t_1},\mathbf{a}_\sigma^{t_2})(0)=0.
\end{equation}
\end{IEEEproof}
The proposed SZCCS in \textit{Theorem} \ref{thm5.1}, is optimal since $K=p^{k+\delta}=p^k(p^m/p^{m-\delta})=M(N/(Z+1))$.
\begin{remark} \label{rem1}
For $\delta=0$, the proposed construction generates $(p^k,p^k,p^m)$-CCC. So, the construction of $(p^k,p^k,p^m)$-CCC, in \cite{sarkar,nishant_cczcz} become  particular cases of the proposed construction.
\end{remark}
\begin{remark} \label{rem2}
For $p=2$, the proposed construction generates $(2^{k+\delta},2^k,2^m,2^{m-\delta})$-ZCCS, so the available construction of ZCCS provided in \cite{wuyu} is a special case of the proposed construction.
\end{remark}
\begin{remark} \label{rem3}
In \cite{zccs_ccds_2022}, direct construction of $(p^{\delta+1},p,p^m,p^{m-\delta})$-ZCCS is provided, the proposed construction with $k=1$ generates the ZCCS with the same parameter.
\end{remark}
\begin{remark} \label{rem4}
Since every SZCCS is a ZCCS, the construction of $(p^{n+v},p^n,p^m,p^{m-v})$-ZCCS in \cite{multiple_ccc_2022} occurs as a special case of the proposed construction in \textit{Theorem} \ref{thm5.1}.
\end{remark}
\begin{remark} \label{rem5}
The construction of $(8,2,2^m,2^{m-2}-1)$-SZCCS is given is \cite{szccs1}, which appears as a special case of the proposed construction when $p=2,\delta=2,k=1.$
\end{remark}
We provide the following example to illustrate how optimal SZCCS is obtained from \textit{Theorem} \ref{thm5.1}.
\begin{example}\label{ex5.1}
For $m=3,\delta=1$, let $\pi_1=\pi$ is the identity permutation of $\{1,2\}$, i.e. $\pi(1)=1$ and $\pi(2)=2$. Further let us take $p=3$ and $q=3$ and define the MVF $g:\{0,1,2\}^3 \rightarrow \mathbb{Z}_3$ as
  $g(x_1,x_2,x_3)=x_1x_2$. Also let us define the sets $A^t=\{\mathbf{f}+\sigma_1x_1+t_1x_2+t_2x_3:\sigma_1 \in \{0,1,2\}\}$, for $t=0,1,\hdots,8$ with $p$-ary representation $(t_1,t_2)$. So, from \textit{Theorem} \ref{thm5.1}, $\mathrm{A}=\{A^0,A^1,\hdots,A^8\}$ is a $(9,3,27,8)$-SZCCS. The codes $A^0,A^1,\hdots,A^8$ are listed in Table \ref{table 5.1} explicitly, where by integer $i$, we mean $\omega^i$ where $\omega=exp(2\pi\sqrt{-1}/3)$, and AACF graph of $A^0$ is plotted in Fig. \ref{fig5.1}, and ACCF graph of $A^2$ and $A^8$ in Fig. \ref{fig5.2}.
\end{example}
\begin{table}[h]
\centering
\caption{Different codes of the SZCSS obtained from \textit{Example} \ref{ex5.1}}
\resizebox{\textwidth}{!}{
\begin{tabular}{ | c | c| } 
\hline
    $A^0$ &  $A^1$  \\ \hline
  $  \begin{matrix}
     0     0     0     0     1     2     0     2     1     0     0     0     0     1     2     0     2     1     0     0  0     0     1     2     0     2     1 \\
      0     0     0     1     2     0     2     1     0     0     0     0     1     2     0     2     1     0     0     0  0     1     2     0     2     1     0 \\
   0     0     0     2     0     1     1     0     2     0     0     0     2     0     1     1     0     2     0     0  0     2     0     1     1     0     2 
    \end{matrix}$
 &  
$ \begin{matrix}
  0     1     2     0     2     1     0     0     0     0     1     2     0     2     1     0     0     0     0     1   2     0     2     1     0     0     0 \\
  0     1     2     1     0     2     2     2     2     0     1     2     1     0     2     2     2     2     0     1   2     1     0     2     2     2     2 \\
  0     1     2     2     1     0     1     1     1     0     1     2     2     1     0     1     1     1     0     1   2     2     1     0     1     1     1
 \end{matrix}$
   \\ \hline 
$A^2$  & $A^3$   \\ \hline
$\begin{matrix}
 0     2     1     0     0     0     0     1     2     0     2     1     0     0     0     0     1     2     0     2   1     0     0     0     0     1     2  \\
 0     2     1     1     1     1     2     0     1     0     2     1     1     1     1     2     0     1     0     2   1     1     1     1     2     0     1  \\
   0     2     1     2     2     2     1     2     0     0     2     1     2     2     2     1     2     0     0     2  1     2     2     2     1     2     0
\end{matrix}
$
&
$\begin{matrix}
 0     0     0     0     1     2     0     2     1     1     1     1     1     2     0     1     0     2     2     2  2     2     0     1     2     1     0  \\
   0     0     0     1     2     0     2     1     0     1     1     1     2     0     1     0     2     1     2     2   2     0     1     2     1     0     2  \\
    0     0     0     2     0     1     1     0     2     1     1     1     0     1     2     2     1     0     2     2    2     1     2     0     0     2     1
\end{matrix}$
\\ \hline 
$A^4$  &  $A^5$  \\ \hline
$\begin{matrix}
0     1     2     0     2     1     0     0     0     1     2     0     1     0     2     1     1     1     2     0  1     2     1     0     2     2     2  \\
 0     1     2     1     0     2     2     2     2     1     2     0     2     1     0     0     0     0     2     0   1     0     2     1     1     1     1  \\
   0     1     2     2     1     0     1     1     1     1     2     0     0     2     1     2     2     2     2     0  1     1     0     2     0     0     0
\end{matrix}$
&
$\begin{matrix}
0     2     1     0     0     0     0     1     2     1     0     2     1     1     1     1     2     0     2     1    0     2     2     2     2     0     1   \\
 0     2     1     1     1     1     2     0     1     1     0     2     2     2     2     0     1     2     2     1     0     0     0     0     1     2     0  \\
 0     2     1     2     2     2     1     2     0     1     0     2     0     0     0     2     0     1     2     1   0     1     1     1     0     1     2
\end{matrix}$
\\ \hline 
$A^6$  &  $A^7$  \\ \hline
$\begin{matrix}
0     0     0     0     1     2     0     2     1     2     2     2     2     0     1     2     1     0     1     1   1     1     2     0     1     0     2  \\
   0     0     0     1     2     0     2     1     0     2     2     2     0     1     2     1     0     2     1     1  1     2     0     1     0     2     1  \\
   0     0     0     2     0     1     1     0     2     2     2     2     1     2     0     0     2     1     1     1   1     0     1     2     2     1     0
\end{matrix}$
&
$\begin{matrix}
 0     1     2     0     2     1     0     0     0     2     0     1     2     1     0     2     2     2     1     2    0     1     0     2     1     1     1  \\ 
  0     1     2     1     0     2     2     2     2     2     0     1     0     2     1     1     1     1     1     2    0     2     1     0     0     0     0  \\
  0     1     2     2     1     0     1     1     1     2     0     1     1     0     2     0     0     0     1     2   0     0     2     1     2     2     2
\end{matrix}$ \\ \hline
\multicolumn{2}{|c|}{
$A^8$ }  \\ \hline 
\multicolumn{2}{|c|}{
$\begin{matrix}
0     2     1     0     0     0     0     1     2     2     1     0     2     2     2     2     0     1     1     0    2     1     1     1     1     2     0   \\
 0     2     1     1     1     1     2     0     1     2     1     0     0     0     0     1     2     0     1     0     2     2     2     2     0     1     2  \\
  0     2     1     2     2     2     1     2     0     2     1     0     1     1     1     0     1     2     1     0   2     0     0     0     2     0     1
\end{matrix}$ }\\  \hline
\end{tabular}}\label{table 5.1}
\end{table}
\begin{figure}[h]
    \centering
    \includegraphics[width=9cm, height=5cm]{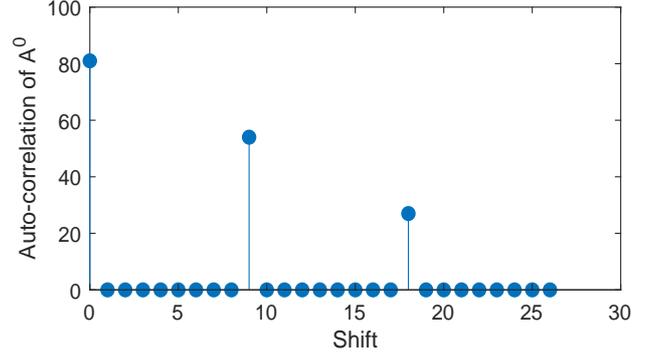}
    \caption{AACF of $A^0$.}
    \label{fig5.1}
\end{figure}
\begin{figure}[h]
    \centering
    \includegraphics[width=9cm, height=5cm]{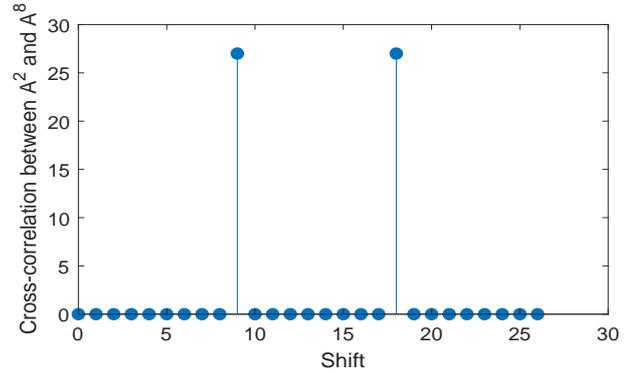}
    \caption{ACCF of $A^2$ and $A^8$.}
    \label{fig5.2}
\end{figure}
\section{Comparison With Existing Works}

In the literature, research on SZCCS is relatively recent, with only constructions available for $(8,2,2^{m},2^{m-2}-1)$-SZCCS \cite{szccs1} and $(2,2,2^{m-1}+2^v,2^v-1)$-SZCCS \cite{szccs2}. Optimal SZCCS are utilized in the design of optimal training sequences for GSM systems, which can achieve superior channel estimation performance compared to other sequence classes \cite{szccs1,szccs2}. The proposed construction in \textit{Theorem} \ref{thm5.1} produces optimal SZCCS with variable set size $(p^{k+\delta})$, flock size $(p^k)$, sequence length $(p^m)$, and ZCZ width $(p^{m-\delta}-1)$. Thus, due to the flexibility of parameters, the proposed SZCCS offers adaptability in generating optimal training matrices for GSM systems with multiple active transmit antennas \cite{szccs2}.



Based on definitions \ref{def2} and \ref{DEF3}, it is clear that a ZCCS can be regarded as a particular case of an SZCCS, where $\mathcal{T}_2=\phi$ and $\mathcal{T}_1=\{1,2,\hdots, Z-1\}$. Moreover, according to \textit{Lemma} \ref{lem5.1}, optimality of SZCCS implies optimality of ZCCS. Consequently, as detailed in \textit{Remark} \ref{rem1}-\ref{rem4}, the existing constructions of CCC in \cite{sarkar,nishant_cczcz} and ZCCS in \cite{wuyu,zccs_ccds_2022,multiple_ccc_2022} can be viewed as special cases of the proposed construction.

\section{Conclusion}
An optimal SZCCS of prime power length has been constructed directly using MVF in this paper. Since SZCCS have both front-end and tail-end ZCZ width, they are used in designing optimal training sequences for broadband GSM systems over frequency-selective channels. The proposed MVF-based construction generates optimal $(p^{k+\delta},p^k,p^m,p^{m-\delta}-1)$-SZCCS, which generalizes many of the existing works of ZCCS and SZCCS. 

\bibliographystyle{IEEEtran}
\bibliography{reference}

\end{document}